\documentclass{optica-article}

\journal{opticajournal} % for journals or Optica Open

\articletype{Research Article}

\usepackage{lineno}
\usepackage{subcaption}
\usepackage{siunitx}
\begin{document}

\title{Transport-of-Intensity Model for Single-Mask X-ray Differential Phase Contrast Imaging}

\author{Jingcheng Yuan\authormark{1} and Mini Das\authormark{1,2*}}

\address{\authormark{1}Department of Physics, University of Houston, 3507 Cullen Blvd, Houston, TX 77204, USA\\
\authormark{2}Department of Electrical and Computer Engineering, Department of Biomedical Engineering, University of Houston, 3517 Cullen Blvd, Houston, TX 77204, USA}

\email{\authormark{*}mdas@uh.edu} %% email address is required; see note below about the corresponding author designation
%{asbstract*}
% use {asbstract*} to suppress the copyright line. Copyright information will be added in production

\begin{abstract*} 
\textcolor{black}{X-ray phase contrast imaging holds great promise for improving the visibility of light-element materials such as soft tissues and tumors. Single-mask differential phase contrast imaging method stands out as a simple and effective approach to yield differential phase contrast. In this work, we introduce a novel model for a single-mask phase imaging system based on the transport-of-intensity equation. Our model provides an accessible understanding of signal and contrast formation in single-mask X-ray phase imaging, offering a clear perspective on the image formation process, for example, the origin of alternate bright and dark fringes in phase contrast intensity images. Aided by our model, we present an efficient retrieval method that yields differential phase contrast imagery in a single acquisition step. Our model gives insight into the contrast generation and its dependence on the system geometry and imaging parameters in both the initial intensity image as well as in retrieved images. The model validity as well as the proposed retrieval method is demonstrated via both experimental results on a system developed in-house as well as with Monte Carlo simulations. In conclusion, our work not only provides a model for an intuitive visualization of image formation but also offers a method to optimize differential phase imaging setups, holding tremendous promise for advancing medical diagnostics and other applications.}
\end{abstract*}

%%%%%%%%%%%%%%%%%%%%%%%%%%  body  %%%%%%%%%%%%%%%%%%%%%%%%%%
\section{Introduction}
Conventional X-ray imaging relies on the variations of X-ray attenuation properties among different tissue types. However, it has limited contrast for low atomic number materials such as organs, tumors, and other soft tissue\cite{lewisMedicalPhaseContrast2004}\cite{fitzgeraldPhaseSensitiveRay2000}\cite{taoPrinciplesDifferentXray2021}\cite{auweterXrayPhasecontrastImaging2014}. In recent years, X-ray phase contrast imaging (PCI) has gained much attention for its potential to enhance this soft tissue contrast by utilizing relative phase changes with X-ray propagation through the object. Among the various techniques available, single-mask differential phase contrast imaging method stands out as a simple and effective approach yielding higher contrast than optic-free methods like propagation based phase imaging. \textcolor{black}{Other single-optical element methods, such as the speckle tracking technique, also yield favorable outcomes. However, they necessitate the use of high-resolution detectors and may potentially demand a higher X-ray dose.\cite{morganXrayPhaseImaging2012}\cite{zhouSpecklebasedXrayPhasecontrast2015}}\\
The simple propagation-based (PB) phase contrast imaging, does not require any additional optics in the beam path, but only an increase in the object-to-detector distance and a partially coherent source \cite{wilkinsPhasecontrastImagingUsing1996}\cite{suzukiXrayRefractionenhancedImaging2002}. At a longer propagation distance, the wavefront distortions caused by the object are recorded as intensity variations on the detector plane. These variations can be modeled by the approximated form of the transport-of-intensity equation (TIE)\cite{zuoTransportIntensityEquation2020}:
\begin{equation}
    I(z,\vec{r}) = I(0,\vec{r}) - \frac{z}{k} (\nabla_{\perp}I(0,\vec{r})\cdot\nabla_{\perp}\phi(\vec{r})+ I(0,\vec{r}) \nabla_{\perp}^{2}\phi(\vec{r})) 
    \label{Eqn_TIE}
\end{equation}
Here $I(z,\vec{r})$ and $I(0,\vec{r})$ are the x-ray intensity at the object plane and detector plane respectively, $\phi(\vec{r})$ is the beam's phase shift caused by the object, $z$ is the object-to-detector distance, $k$ is the wave number, and $\vec{r}$ is the coordinate in x-y plane. In most applications of interest with predominantly soft materials in the beam path, we can assume the intensity variation is slow in the x and y direction, so the second term $\nabla_{\perp}I(0,\vec{r})\cdot\nabla_{\perp}\phi(\vec{r})$ can be neglected\cite{zuoTransportIntensityEquation2020}\cite{paganinSimultaneousPhaseAmplitude2002}. Hence the equation becomes:
\begin{equation}
	I(z,\vec{r}) = I(0,\vec{r}) - \frac{z}{k} I(0,\vec{r}) \nabla_{\perp}^{2}\phi(\vec{r})
	\label{Eqn_TIE_FSP}
\end{equation}
\\Thus in addition to the attenuation signal ($I(0,\vec{r})$), the intensity at each detector pixel is predominantly influenced by the Laplacian of X-ray phase shift caused by the object. This Laplacian phase signal manifests as bright and dark borders along the edges, leading to edge enhancement. % (Fig. \ref{Fig_Propagation})
\begin{figure}[ht]
	\centering
	\begin{subfigure}{0.42\textwidth}
		\includegraphics[width=\textwidth]{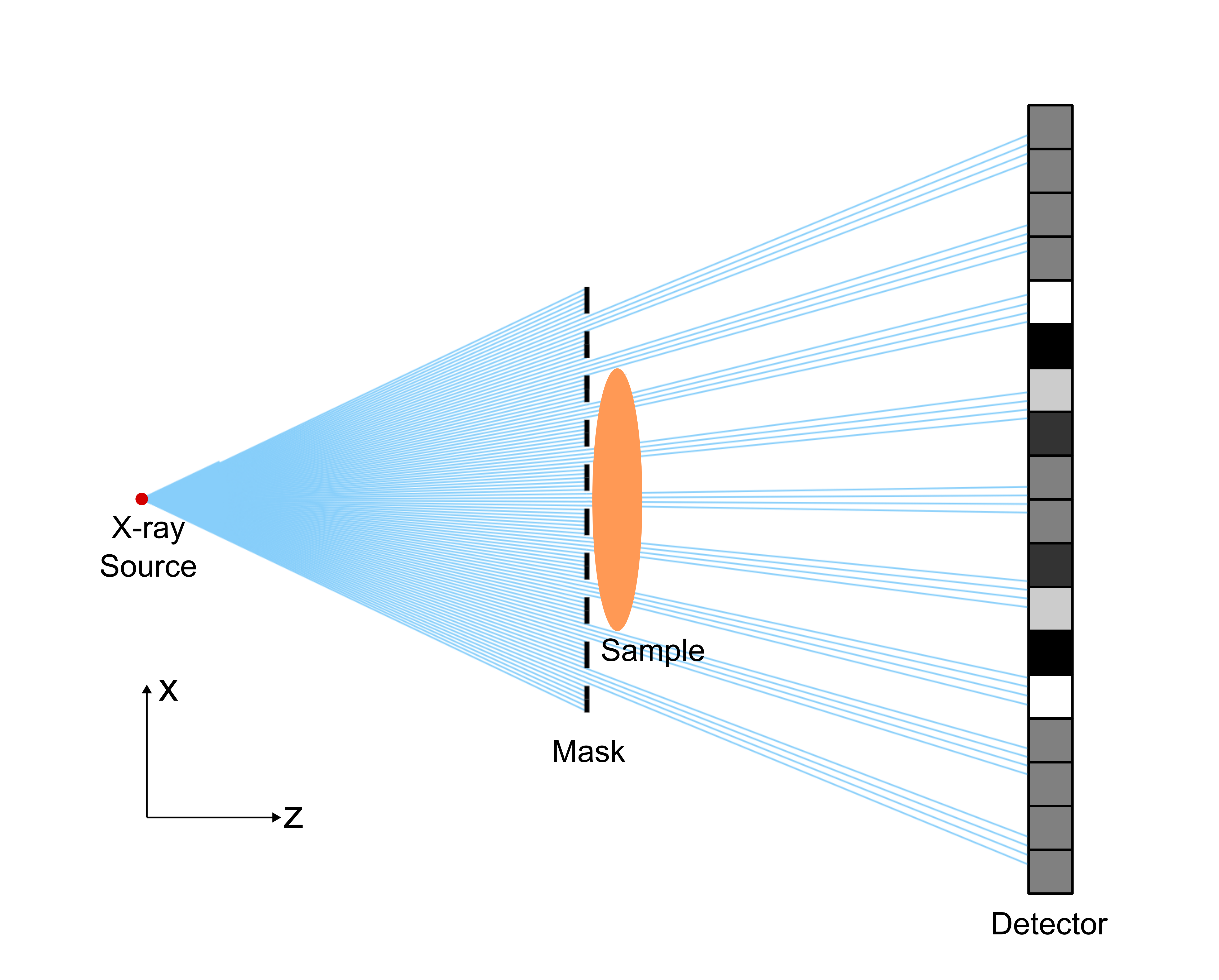}
		\caption{} \label{Fig_SM_SetUp}
	\end{subfigure}
	\hfil
	\begin{subfigure}{0.42\textwidth}
		\includegraphics[width=\textwidth]{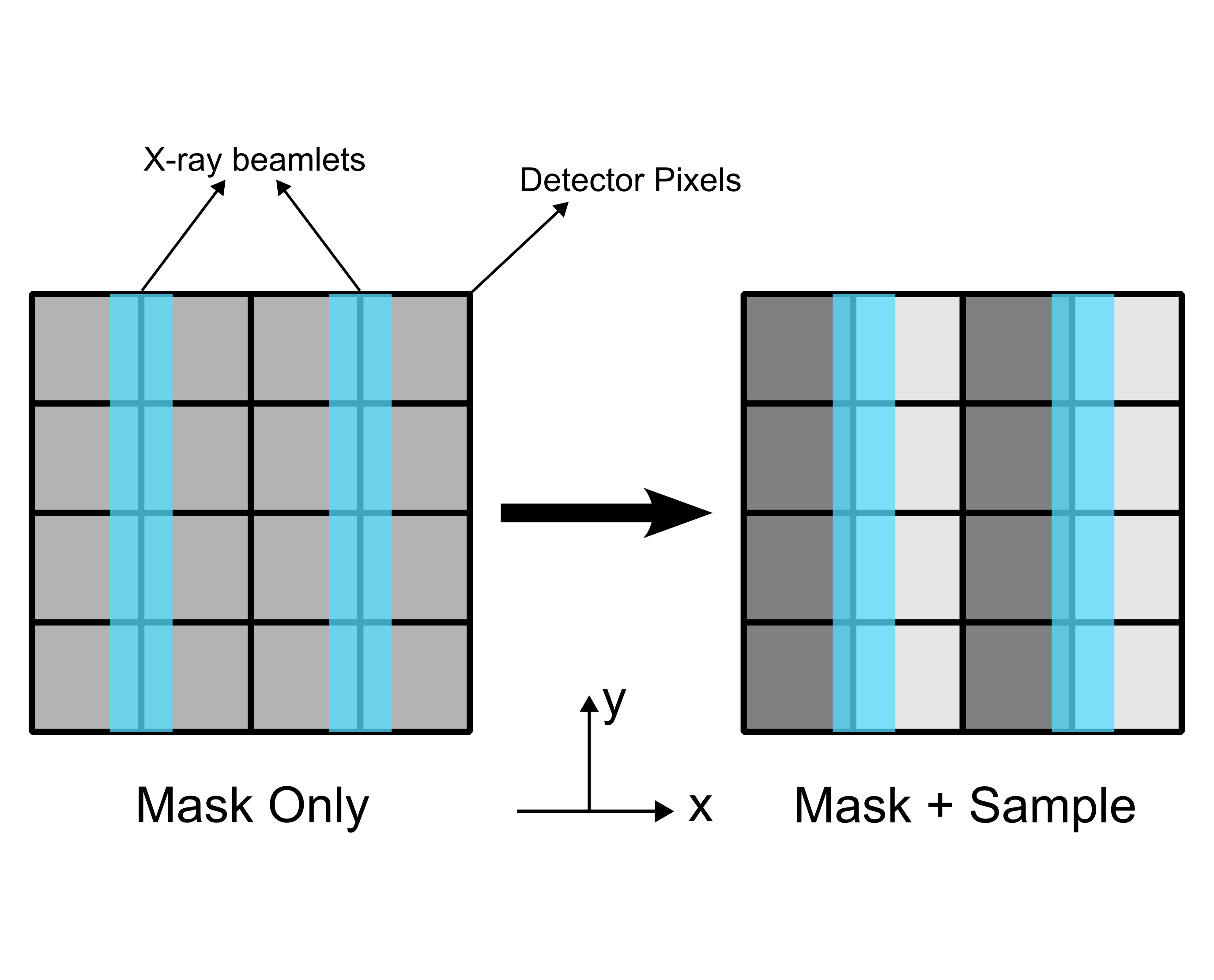}
		\caption{} \label{Fig_SM_Pixels}
	\end{subfigure}
	\caption{Schematic of the single mask phase imaging method. (a) Top view of the set up. The X-ray beam propagates in $z$ direction, and the detector pixels are placed in the x-y plane. (b) Diagram of mask alignment with detector pixels. The mask strips are along y direction.}\label{Fig_Diagram}
\end{figure}
\\A single-mask phase imaging technique \cite{krejciHardXrayPhase2010} is similar to PB phase imaging but with an added periodic X-ray absorption mask positioned between the source and the object, in close proximity to the object (Fig. \ref{Fig_SM_SetUp}). The mask creates X-ray beamlets by periodically blocking X-rays with thin strips of heavy-element materials like gold. The mask is aligned with respect to the detector such that the center of each thin and long strip of beamlet is aligned to every other pixel boundary \cite{krejciHardXrayPhase2010}\cite{kallonComparingSignalIntensity2017}. Hence, with a proper alignment, in the absence of an object in the beam path, the signal intensity on each detector pixel column is uniform, showing no discernible patterns (Fig. \ref{Fig_SM_Pixels})). When the object is introduced, the heterogeneities within the object induce refraction effects that alter the original directions of the beamlets. Thus, intensity differences appear between neighboring pixels, resulting in the appearance of bright and dark fringes on the detector. Fig. \ref{Fig_Diagram}b shows the schematic and \ref{Fig_SM_Raw} shows experimental results to be described in detail later. These relative intensity variations can allow disentangling differential phase information from attenuation-related intensity variations on the detector plane when the appropriate light-transport model is known.\\
\textcolor{black}{In terms of wave optics, this can also be explained as the modification of the Fresnel diffraction pattern of the periodic mask with the introduction of the object. However, given the X-ray wavelength and the mask geometry, the resolution of fine-structure induced by diffraction effect is at sub-micrometer scale, a dimension considerably smaller than the detector pixel size. Moreover, these fine structure features from mask diffraction also get blunted due to any focal spot width of the source. Consequently, these fine structures from diffraction have minimal contribution to the signal measured by the detector. Thus, for the relatively low resolution detectors (tens of micrometers pixel sizes), used for the current X-ray imaging systems, with an initial mask alignment with the detector as shown in Fig. 1b, a uniform illumination is observed. The shift in Fresnel diffraction patterns with the introduction of the object yields intensity variations shown as bright and dark fringes which will be explained also by the TIE model derived in this paper. Here again, one does not capture the fine details of the diffraction patterns, but rather the shift in intensity with and without the object which can be easily captured by even the lower resolution detectors.}\\
We note that the single-mask PCI method is a significantly simplified version of the double-mask edge-illumination (EI) method developed earlier \cite{olivoCodedapertureTechniqueAllowing2007}\cite{olivoNoninterferometricPhasecontrastImages2011} and also avoids 'wasting' large number of photons that has already transmitted through the object.
\begin{figure}[ht]
	\centering
	\begin{subfigure}{0.42\textwidth}
		\includegraphics[height=0.19\textheight]{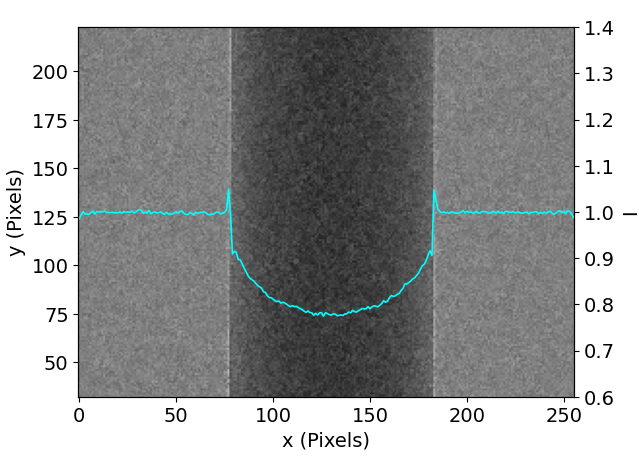}
		\caption{} \label{Fig_Propagation}
	\end{subfigure}
	\begin{subfigure}{0.56\textwidth}
		\includegraphics[height=0.19\textheight]{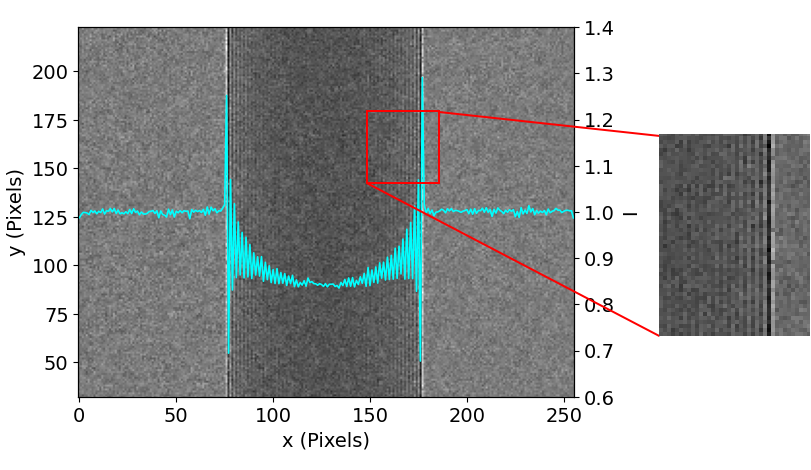}
		\caption{} \label{Fig_SM_Raw}
	\end{subfigure}
	\caption{PCI intensity images of a PMMA rod and their cross-section profile (blue curve) with (a) propagation-based method and (b) single-mask method. \textcolor{black}{The y-axis ticks on the left indicate positions in pixels, while the ticks on the right correspond to the cross-section profile curve.}}
\end{figure}
\\The formulation of single-mask PCI has been previously attempted using both refraction\cite{krejciHardXrayPhase2010} and wave-optics\cite{vittoriaStrategiesEfficientFast2013} models. However, these existing models have limitations in terms of providing intuitive visualizations of signal and contrast formation in the images. In this paper, we present a new model based on TIE and show how this model can be used for efficient retrieval of absorption and differential phase. Furthermore, we show a single-shot (only one acquisition with no movement of object or optical components), low-dose phase-imaging that yields multiple image features and contrast types. While our prior work has shown efficient phase retrieval methods with spectral data (using photon counting detectors) \cite{gursoySinglestepAbsorptionPhase2013}\cite{dasSingleStepXray2016}\cite{dasSpectralXrayPhase2014}\cite{dasSingleStepDifferential2016}\cite{vazquezQuantitativePhaseRetrieval2021}, the retrieval shown here does not require spectral data.

\section{Methods}
\subsection{Formulation}
Our formulation for single-mask PCI starts with the TIE, (Eq. (\ref{Eqn_TIE})). Unlike the propagation-based method, here we have a high-contrast periodic absorption mask, so the term $\nabla_{\perp}I(0,\vec{r})\cdot\nabla_{\perp}\phi(\vec{r})$ can no longer be neglected. Here, the transmitted intensity at the object plane is $I(0,\vec{r})=T(\vec{r})\cdot M(x)$, where $T(\vec{r})$ and $M(x)$ is the transmission function of the object and the mask, respectively. Therefore:
\begin{equation}
	 \nabla_\perp I(0,\vec{r}) = T\nabla_\perp M + M\nabla_\perp T\approx T\partial_x M
	 \label{Eqn_Grad_I}
\end{equation}
Here we applied the approximation that $\nabla_\perp I(0,\vec{r})$ is mainly contributed by the mask so $M\nabla_\perp T$ can be neglected. After substituting Eq. (\ref{Eqn_Grad_I}) into Eq. (\ref{Eqn_TIE}), the x-ray intensity measured by each detector pixel can be calculated by integrating Eq. (\ref{Eqn_TIE}) over the range of the corresponding pixel:
\begin{equation}
    \begin{aligned}
        I_n&= \int_{x_n}^{x_{n+1}}T\cdot M\,\mathrm{d}x -\frac{z}{k}\int_{x_n}^{x_{n+1}}T\cdot\partial_x M\cdot\partial_x\phi\,\mathrm{d}x-\frac{z}{k}\int_{x_n}^{x_{n+1}}T\cdot M\cdot\nabla_\perp^2 \phi\,\mathrm{d}x\\
        \label{Eqn_SM_Integrate1}
    \end{aligned}
\end{equation}
where $n$ is the pixel index in the horizontal direction when the masks with slits in the vertical direction is used, $x_n$ and $x_{n+1}$ is the coordinate of the left and right boundary of the corresponding pixel respectively.
\\Here we assume that the attenuation, phase, and differential phase of the sample vary slowly within the range of a pixel. Then Eq. (\ref{Eqn_SM_Integrate1}) becomes:
\begin{equation}
    \begin{aligned}
        I_n&= T_n \int_{x_n}^{x_{n+1}}M(x)\,\mathrm{d}x -\frac{z}{k} T_n\partial_x\phi_n\int_{x_n}^{x_{n+1}}\partial_x M(x)\,\mathrm{d}x -\frac{z}{k} T_n \nabla_\perp^2 \phi_n \int_{x_n}^{x_{n+1}}M(x)\,\mathrm{d}x \\
        &= T_n(1-L_n)\int_{n p}^{(n+1)p}M(x)\,\mathrm{d}x - T_n D_n \int_{n p}^{(n+1)p}\partial_x M(x)\,\mathrm{d}x 
        \label{Eqn_SM_Integrate2}
    \end{aligned}
\end{equation}
Here $T_n = T(x_n)$, which represents the object attenuation function averaged within each pixel; $L_n = \frac{z}{k}\nabla^2_\perp\phi(x_n)$, which is the Laplacian of phase shift caused by the object; $D_n = \frac{z}{k}\partial_x\phi(x_n)$, which is the gradient of phase shift, and is proportional to the x-ray refraction angle. 
\\In the case of a perfect mask, as we have demonstrated in our previous paper \cite{dasSpectralXrayPhase2014}\cite{dasApproximatedTransportofintensityEquation2014}, the mask transmission function $M(x)$ can be expressed as a square wave. Considering the imperfection of the mask, a more general form of its transmission function can be expressed as a Fourier series:
\begin{equation}
	\begin{aligned}
		M(x) =  \sum_{m} C_m\cos{\frac{2\pi m x}{(2p)}}\\
	\end{aligned}
	\label{Eqn_Mask_Fourier}
\end{equation}
where $p$ is the detector pixel size, which means the period of the mask is two times of detector pixel size (See Fig. \ref{Fig_SM_SetUp} for reference on mask vs detector period). Then we have: 
\begin{subequations}
	\begin{align}
		&\int_{n p}^{(n+1)p}M(x)\,\mathrm{d}x =  \int_{n p}^{(n+1)p}\sum_{m} C_m\cos{\frac{\pi m x}{p}}\,\mathrm{d}x 
		=C_0 p\label{Eqn_Mask_1}\\ 
		&\int_{n p}^{(n+1)p}\partial_x M(x)\,\mathrm{d}x =  M((n+1)p) - M(np)  
		=-2 \sum_{m}C_{2m+1}\cdot(-1)^n \label{Eqn_Mask_2}
	\end{align}\label{Eqn_Mask}
\end{subequations}
The results of Eq. (\ref{Eqn_Mask}) are related to the mask transmission function and are not related to the object property. Thus, Eq. (\ref{Eqn_SM_Integrate2}) becomes:
\begin{equation}
    I_n= w_e T_n(1-L_n) - \alpha (-1)^n T_n D_n 
    \label{Eqn_SM_Final}
\end{equation}
One can see from this equation that the signal is a combination of two distinct effects. The first term in Eq. (\ref{Eqn_SM_Final}), which we refer to as the propagation-based (PB) part, shares the same form as the propagation-based PCI (Eq. (\ref{Eqn_TIE_FSP})). 
\\The second term, referred to as the differential phase contrast (DPC) term, gives rise to the characteristic bright and dark fringes within the image, as demonstrated in the example depicted in Fig. \ref{Fig_SM_Raw}. This is because it contains the factor $(-1)^n$, where $n$ denotes the pixel column index. The magnitude of these fringes is directly proportional to the DPC signal $D_n$.
\\Also, the two parts of the signal are multiplied by two mask-related coefficients $w_e$ and $\alpha$ respectively. The coefficients' values are determined by combining Eq. (\ref{Eqn_SM_Integrate2}) and Eq. (\ref{Eqn_Mask}).
\begin{subequations}
    \begin{align}
        w_e &= C_0 p \label{Eqn_SM_w}\\
        \alpha &= 2 \sum_{m}C_{2m+1} \label{Eqn_SM_alpha}
    \end{align}
\end{subequations}
\\According to Eq. (\ref{Eqn_Mask_1}), the coefficient $w_e$ represents the integration of the mask-transmission function within a pixel, which corresponds to the average transmission of the mask. Thus, it can be interpreted as the effective transparent width or aperture size. It is also similar to the $w_e$ in our previous model for double mask method\cite{dasApproximatedTransportofintensityEquation2014}. On the other hand, $\alpha$ is a unit-less coefficient that depends on the odd Fourier coefficients of the mask's transmission function. It can be understood as \textcolor{black}{the attenuation contrast between the blocked and transmitted region of the mask.}
\\The two coefficients can be interpreted as separate filters for the PB part and the DPC part independently. In comparison with the PB method, the intensity of the PB signal in the single-mask method is reduced by the coefficient $w_e$. This implies that the mask selectively reduces the X-ray intensity that contributes to the PB part, thereby allowing for a reduction in X-ray radiation dose to the sample without affecting the signal intensity of the DPC part. The second coefficient, $\alpha$, which represents the contrast of the mask, determines the efficiency of obtaining the DPC signal.

\subsection{Retrieval Method}
From the last section we could see the attenuation, Laplacian phase and differential phase have contributions to the measured intensity. Among them, the DPC part is shown as high-frequency fringes in Fig. \ref{Fig_SM_Raw}. A retrieval process is needed to separate the PB part and the DPC part.\\
In an experimental realization, a single image is taken with the object and the mask in the beam path. This image (represented as $I_{n(M+S)}$) can be compared with the image with mask only (flat field $I_{n(M)}$). The formula for the mask-and-sample image $I_{n(M+S)}$ is shown in Eq. (\ref{Eqn_SM_Final}); for the mask-only (or flat-field) image, $I_{n(M)} = w_e$. After doing flat-field correction, we obtain:
\begin{equation}
    \bar{I}_{n} = \frac{I_{n(M+S)}}{I_{n(M)}} = T_n(1-L_n)- (-1)^n\frac{\alpha}{w_e}T_n D_n 
    \label{Eqn_SM_FF}
\end{equation}
Thus we can write the corrected intensity for $n^{th}$ and $(n+1)^{th}$ pixels in a same row:
\begin{equation}
    \begin{aligned}
        \bar{I}_n &= T_n(1-L_n)- (-1)^n\frac{\alpha}{w_e}T_n D_n \\
        \bar{I}_{n+1} &= T_{n+1}(1-L_{n+1})+ (-1)^n\frac{\alpha}{w_e}T_{n+1} D_{n+1}
    \end{aligned}
\end{equation}
We can separate PB and DPC signals by adding and subtracting the intensity values on $n^{th}$ and $(n+1)^{th}$ pixels \textcolor{black}{in each row}:
\begin{subequations}
    \begin{align}
        \bar{I}_n + \bar{I}_{n+1} &\approx 2 T_n(1-L_n) \label{Eqn_Retrieval_plus}\\
        \bar{I}_n - \bar{I}_{n+1} &\approx 2  (-1)^n\frac{\alpha}{w_e}T_n D_n \label{Eqn_Retrieval_minus}
    \end{align}
\end{subequations}
From Eq. (\ref{Eqn_Retrieval_plus}), we can easily have the retrieval of the PB part (Eq. (\ref{Eqn_SM_FSP})). In order to retrieve $D_n$, if we consider the intensity of the Laplacian of phase to be relatively weak compared with 1, we can apply the approximation of $1-L_n\approx1$ when calculating differential phase $D_n$. Then we can arrive at the retrieval formula for propagation based PCI and differential phase:
\begin{subequations}
    \begin{align}
        T_n(1-L_n) \approx \frac{\bar{I}_n + \bar{I}_{n+1}}{2} \label{Eqn_SM_FSP}\\
        D_n \approx (-1)^n \frac{w_e}{\alpha}\frac{\bar{I}_n-\bar{I}_{n+1}}{\bar{I}_n+\bar{I}_{n+1}} \label{Eqn_SM_DPC}
    \end{align}
\end{subequations}
where (\ref{Eqn_SM_FSP}) is the retrieved PB image and (\ref{Eqn_SM_DPC}) is the retrieved DPC image.
\textcolor{black}{ As one can observe the strength of the retrieved differential phase signal related to the effect slit width $w_e$ and the attenuation contrast of the mask ($\alpha$). }

\subsection{Experiment}
We used a polychromatic micro-focus x-ray tube (Hama-matsu L8121-03) operating with a focal spot of \SI{7}{\micro\metre} and the tube voltage of \SI{40}{\kilo\volt}. The source-to-object and object-to-detector distance were both around \SI{60}{\centi\metre}. The sample in consideration is a PMMA rod with a diameter of \SI{3}{\milli\metre}. We used a mask with gold strips, approximately \SI{52}{\micro\metre} in periodicity, fabricated on a silicon substrate. The data was collected using a Silicon photon-counting detector with the pixel size of \SI{55}{\micro\metre}\cite{ballabrigaMedipix3RXHighResolution2013}, which was carefully calibrated and corrected\cite{dasEnergyCalibrationPhoton2015}\cite{vespucciRobustEnergyCalibration2019}. While spectral data is available with this detector, the methods presented here do not use this spectral information and treats it as an energy integrating detector. The raw image is shown in Fig. \ref{Fig_SM_Raw} and the retrieved PB and DPC images are shown in Fig. \ref{Fig_Experiment}. \textcolor{black}{The results also include images of a multi-material sample shown in Fig. \ref{Fig_Tube}, as well as a dried wasp specimen shown in Fig. \ref{Fig_Wasp}.} 

\section{Results and Discussion}
The raw image (Fig. \ref{Fig_SM_Raw}) obtained from the single-mask method reveals distinct signal components, including attenuation, the Laplacian of phase, and differential phase. The presence of attenuation results in darker regions in the middle of the cylinder. The Laplacian phase manifests as bright and dark borders along the edges. Additionally, the differential phase appears as bright and dark fringes specifically in regions with non-zero phase gradient. As the differential phase signal varies across the sample, it gives rise to variations in the intensity of the fringes. These observed signal components align well with the outcomes predicted by our newly proposed model (Eq. (\ref{Eqn_SM_Final})), validating its reliability in capturing and explaining the underlying physics of the single-mask phase imaging method.\\
\begin{figure}[ht]
	\centering
	\begin{subfigure}{0.49\textwidth}
		\includegraphics[width=\textwidth]{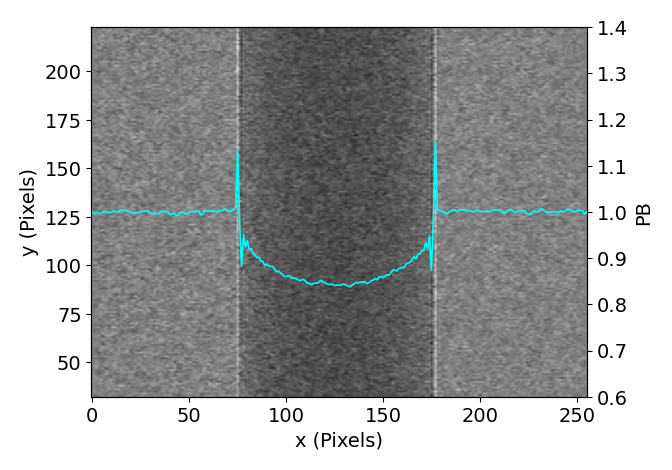}
		\caption{} \label{Fig_SM_PB}
	\end{subfigure}
	\begin{subfigure}{0.49\textwidth}
		\includegraphics[width=\textwidth]{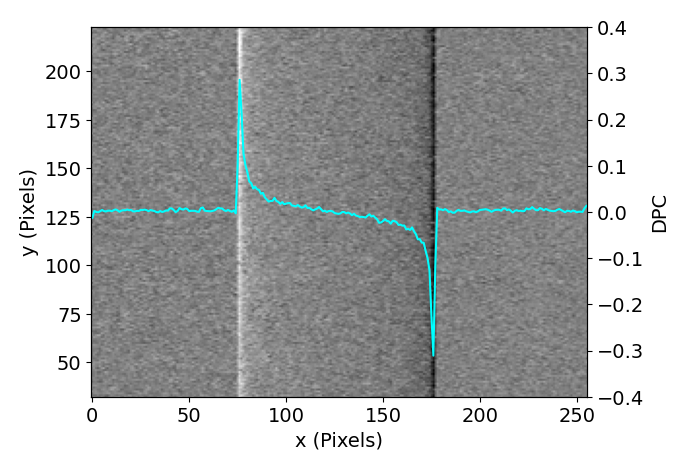}
		\caption{} \label{Fig_SM_DPC}
	\end{subfigure}
	\caption{Retrieved (a) PB image and (b) DPC image of a PMMA rod taken with single mask method in experiment, together with their average cross-section profiles (light blue curve).}\label{Fig_Experiment}
\end{figure}
\\Additionally, The retrieved PB image (Fig. \ref{Fig_SM_PB}) obtained from our model closely resembles the image captured using the propagation-based method (Fig. \ref{Fig_Propagation}). The minor difference between the attenuation levels can be attributed to the shift of the spectrum induced by the mask's silicon substrate. Furthermore, the retrieved differential phase contrast (DPC) image (Fig. \ref{Fig_SM_DPC}) exhibits excellent contrast and visibility. These results indicate that our proposed retrieval method, based on our formulated model, effectively separates and provides visualization of different signal components all from a single image.
\begin{figure}[ht]
	\centering
	\begin{subfigure}{0.26\textwidth}
		\includegraphics[width=\textwidth]{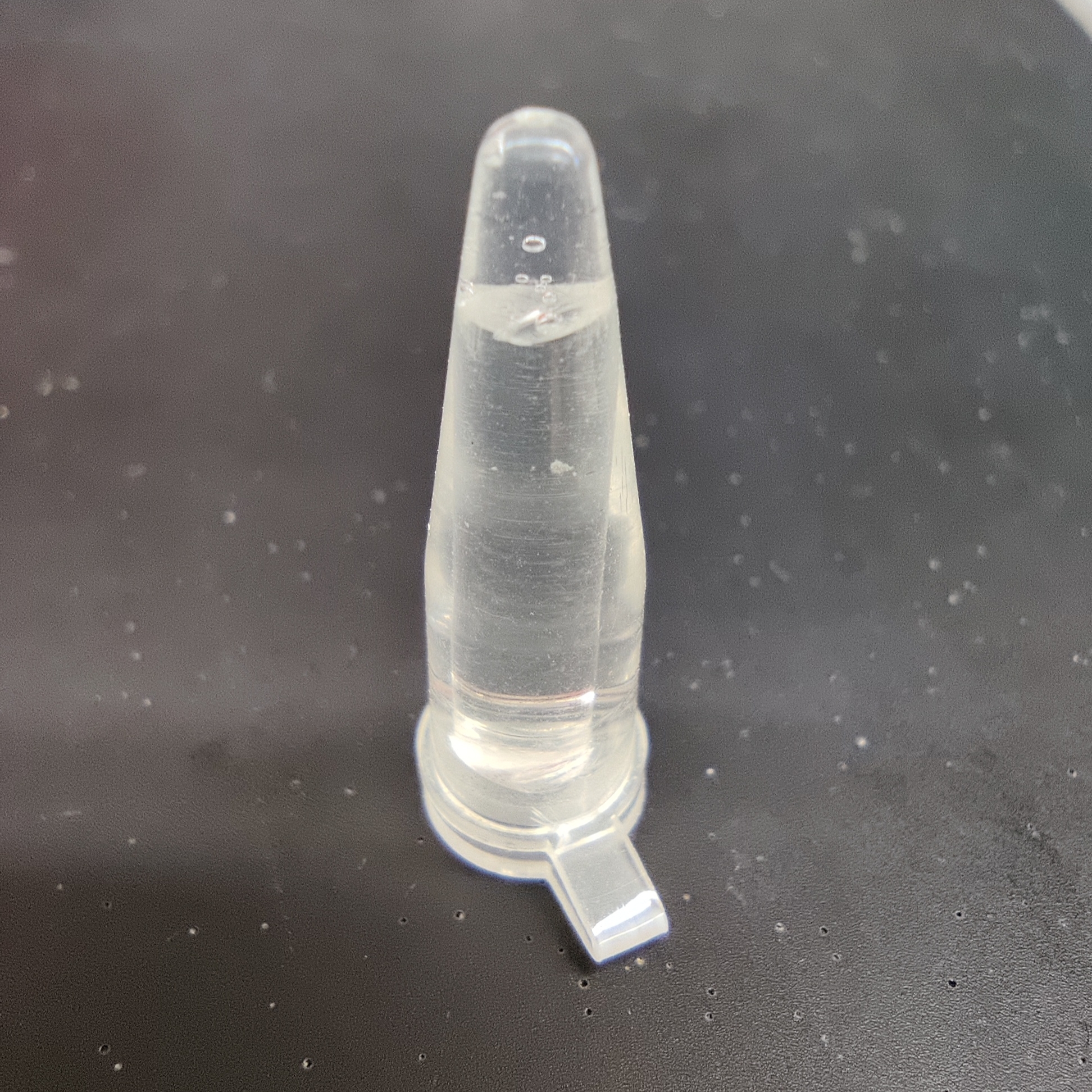}
		\caption{}
	\end{subfigure}
	\begin{subfigure}{0.26\textwidth}
		\includegraphics[width=\textwidth]{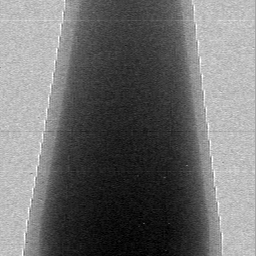}
		\caption{}\label{Fig_Tube_T}
	\end{subfigure}
	\begin{subfigure}{0.26\textwidth}
		\includegraphics[width=\textwidth]{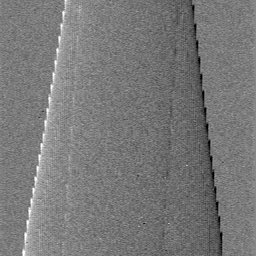}
		\caption{}\label{Fig_Tube_D}
	\end{subfigure}
	\caption{\textcolor{black}{Retrieved images of the middle part of a plastic tube with water and a PMMA rod inside. (a) Picture of the sample; (b) Retrieved PB image; (c) Retrieved DPC image.}}\label{Fig_Tube}
\end{figure}
\textcolor{black}{\\The multi-material sample we used was a plastic tube filled with water and a \SI{3}{\milli\metre} diameter PMMA rod. We could see in PB image (Fig. \ref{Fig_Tube_T}), we can identify the wall of the tube between water but the PMMA rod inside is almost invisible. Conversely, in the DPC image (Fig. \ref{Fig_Tube_D}), the PMMA rod is easily recognizable, although with a lower contrast between the tube wall and water compared to PB image. Both images provide unique and complementary information about the materials within the sample.}
\begin{figure}[ht]
	\centering
	\begin{subfigure}{0.49\textwidth}
		\includegraphics[width=\textwidth]{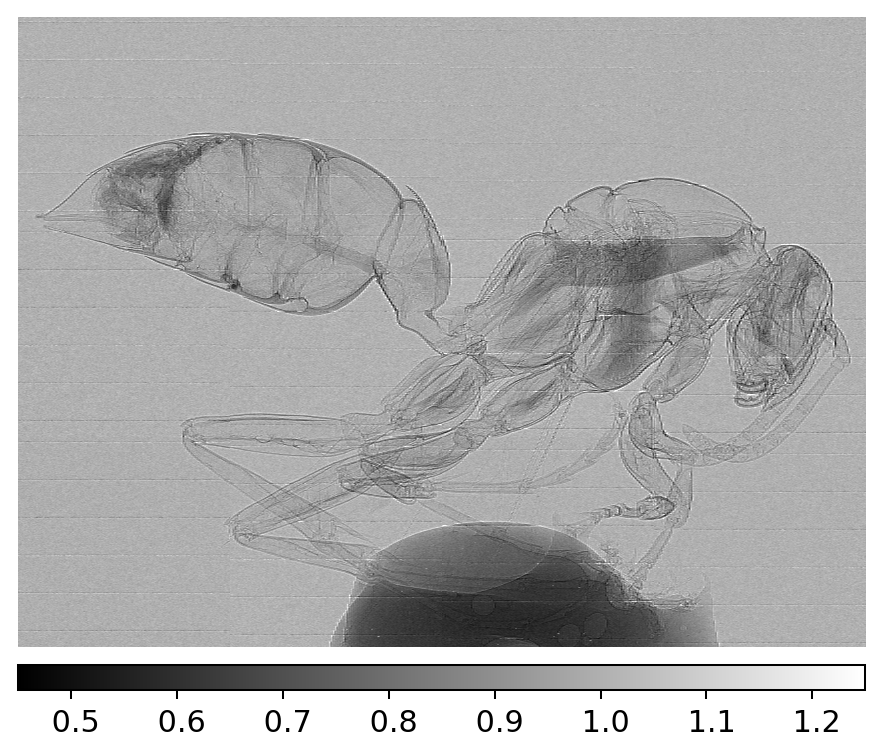}
		\caption{} \label{Fig_Wasp_PB}
	\end{subfigure}
	\begin{subfigure}{0.49\textwidth}
		\includegraphics[width=\textwidth]{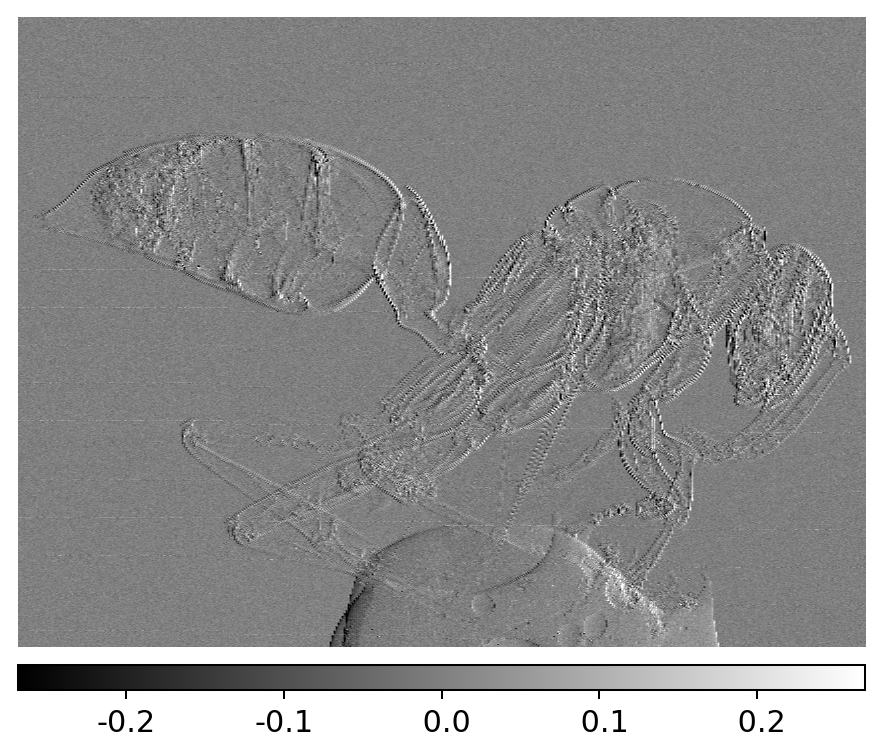}
		\caption{} \label{Fig_Wasp_DPC}
	\end{subfigure}
	\caption{Retrieved images of a wasp specimen taken with single mask method in experiment. (a) Retrieved PB image; (b) Retrieved DPC image.}
	\label{Fig_Wasp}
\end{figure}
\\The retrieved PB and DPC images of the dried wasp using our model are shown in Fig. \ref{Fig_Wasp}. Both these images, retrieved from a single-shot of single-mask phase contrast intensity image, show fine details of the specimen. Both of these images exhibit a high sensitivity to tissue boundaries within the sample, due to the visualization of Laplacian and gradient of phase respectively. Note that the retrieved PB image is a combination of the Laplacian of phase and the attenuation. 
\\Also, discernible differences exist between the two images. The retrieved PB image captures edge information in every orientation in the 2D plane. Conversely, the retrieved DPC image accentuates features that align perpendicular to the mask strips. 
\\In addition, it is interesting to note that the DPC images have higher sensitivity to features with slower variations, such as the bubbles within the adhesive used to affix the specimen shown at the lower region of the image.
\\For further verification of our TIE model, we compared the results obtained from the TIE model and the Monte-Carlo simulation\cite{gierschMonteCarloSimulations2008}. For the TIE model (Eq. (\ref{Eqn_SM_FF})), the calculation is based on known $w_e$ and $\alpha$ without using the mask transmission function. In contrast, the Monte-Carlo simulation employs the mask's transmission function. Both methods model the same experimental geometry and sample. The results are shown in Fig. \ref{Fig_Comparison}. From the figure, we observe that, the results obtained via our TIE model calculation align consistently with the Monte-Carlo simulation outcomes with different mask selections. It shows our new model demonstrates overall accuracy in comparison with Monte-Carlo simulations.
\begin{figure}[ht]
    \centering
    \begin{subfigure}{0.32\textwidth}
    	\centerline{}
    	\centerline{}
    \end{subfigure}
    \begin{subfigure}{0.32\textwidth}
        \centerline{Mask 1}
        \centerline{$w_e=\SI{4.06}{\micro\metre}$, $\alpha=1.0$}
    \end{subfigure}
    \begin{subfigure}{0.32\textwidth}
        \centerline{Mask 2}
        \centerline{$w_e=\SI{13.75}{\micro\metre}$, $\alpha=1.0$}
    \end{subfigure}
    \begin{subfigure}{0.32\textwidth}
        \includegraphics[width=\textwidth]{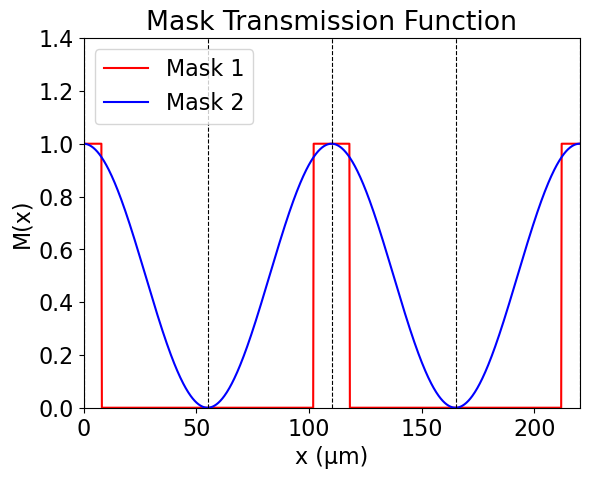}
        \caption{}
    \end{subfigure}
    \begin{subfigure}{0.32\textwidth}
        \includegraphics[width=\textwidth]{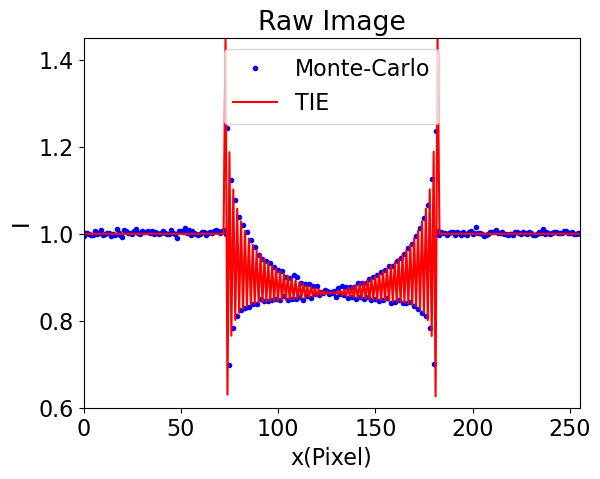}
        \caption{}
    \end{subfigure}
    \begin{subfigure}{0.32\textwidth}
        \includegraphics[width=\textwidth]{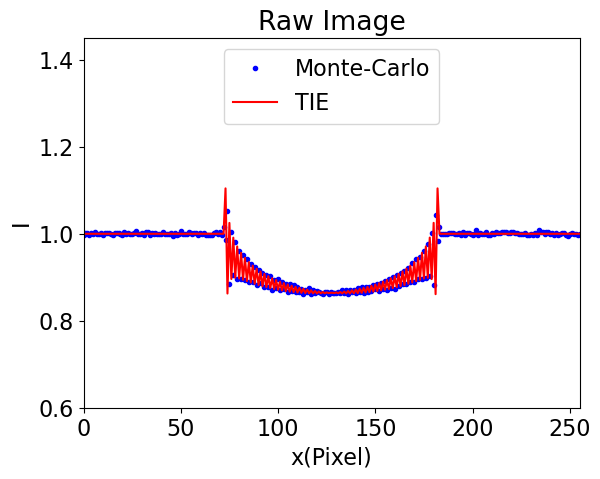}
        \caption{}
    \end{subfigure}
    \caption{Comparison of TIE model calculation and Monte-Carlo simulation with different Mask the transmission functions. (a) Plot of transmission functions, where Mask 1 represents a perfect square wave, while Mask 2 is defined as $M(x)=0.5+0.5\cos(\frac{\pi x}{p})$; (b)-(c) Comparison of the flat-field corrected raw image between TIE model calculation and Monte-Carlo simulation of Mask 1 and Mask 2;}
    \label{Fig_Comparison}
\end{figure}
\\According to our model, the final signal depends not on the specific mask transmission function but rather on two mask parameters: effective aperture size ($w_e$) and mask contrast ($\alpha$). The flat-field corrected raw image in Fig. \ref{Fig_Comparison} reveals components corresponding to Eq. (\ref{Eqn_SM_FF}), including attenuation, DPC, and Laplacian phase. Notably, smaller $w_e$ values yield higher DPC signal contrast, due to increased filtration of photons contributing to the PB term. This enhances DPC signal proportion relative to the PB signal, thus improving X-ray dose efficiency. \textcolor{black}{Furthermore, it is essential to fabricate the mask using heavy element materials such as gold. This choice is driven by the need for sufficient x-ray attenuation while maintaining a relatively small thickness, thereby ensuring a larger contrast parameter ($\alpha$). Also,} it is important to note that smaller $w_e$ values may present challenges in mask manufacturing and potentially require longer exposure times to maintain image quality. Careful consideration of trade-offs between dose efficiency, mask fabrication feasibility, and exposure time is crucial in practical single-mask method applications.

\section{Conclusion}
We have presented a novel light-transport model for single-mask (SM) X-ray phase contrast imaging which yields strong differential phase signatures from a simple system design. The measured X-ray intensity with the SM method combines attenuation, Laplacian phase, and differential phase effects. Our proposed model provides intuitive understanding of the relative contributions of these effects to the detector pixel intensities. Our model also shows how these effects depend on the design parameters of the imaging system. \textcolor{black}{In particular, our newly derived model (Eq. (\ref{Eqn_SM_Final})) gives a clear understanding of the unique bright and dark fringes in phase contrast intensity images.}\\
Aided by our new model, we show an effective retrieval method yielding PB image (combining attenuation with Laplacian phase) and a differential phase contrast image in a single acquisition, thus yielding images with two types of edge enhancement and shape-based contrast. 
\\Our TIE model suggests that the mask transmission function can be characterized by two parameters that have a significant influence on the final signal. By considering these two parameters, one finds a flexibility and adaptability in mask design and performance optimization in practical applications. Our single-shot retrieval method combined with the simple system design yields multiple contrast. This offers a pathway for practical translatability of PCI for a  broad range of applications.

\begin{backmatter}
	
	\bmsection{Funding}
	\textcolor{black}{National Institute of Biomedical Imaging and Bioengineering R01 EB EB029761, DOD CDMRP Breakthrough
	Award BC151607 and National Science Foundation Award1652892.}
	
	\bmsection{Disclosures}
	\textcolor{black}{The authors declare no conflict of interest.}
	
	\bmsection{Data Availability Statement}
	\textcolor{black}{Data underlying the results presented in this paper are not publicly available at this time but may be obtained from the authors upon reasonable request.}
	
\end{backmatter}
%%%%%%%%%%%%%%%%%%%%%%% References %%%%%%%%%%%%%%%%%%%%%%%%%

%Add references with BibTeX or manually.
%\cite{Zhang:14,OPTICA,FORSTER2007,Dean2006,testthesis,Yelin:03,Masajada:13,codeexample}

%%%%%%%%%% If using BibTeX:
\bibliography{JCLibrary.bib}

%%%%%%%%%% If preparing manually:
% \begin{thebibliography}{1}
% \newcommand{\enquote}[1]{``#1''}

% \bibitem{Zhang:14}
% Y.~Zhang, S.~Qiao, L.~Sun, Q.~W. Shi, W.~Huang, L.~Li, and Z.~Yang,
%   \enquote{Photoinduced active terahertz metamaterials with nanostructured
%   vanadium dioxide film deposited by sol-gel method,}
%   {\protect\JournalTitle{Optics Express}} \textbf{22}, 11070--11078 (2014).

% \bibitem{Optica}
% {Optica}, \enquote{{Optica Publishing Group},}
%   \url{http://www.opg.optica.org}.

% \bibitem{FORSTER2007}
% P.~Forster, V.~Ramaswamy, P.~Artaxo, T.~Bernsten, R.~Betts, D.~Fahey,
%   J.~Haywood, J.~Lean, D.~Lowe, G.~Myhre, J.~Nganga, R.~Prinn, G.~Raga,
%   M.~Schulz, and R.~V. Dorland, \enquote{Changes in atmospheric consituents and
%   in radiative forcing,} in \enquote{Climate Change 2007: The Physical Science
%   Basis. Contribution of Working Group 1 to the Fourth assesment report of
%   Intergovernmental Panel on Climate Change,}  S.~Solomon, D.~Qin, M.~Manning,
%   Z.~Chen, M.~Marquis, K.~B. Averyt, M.~Tignor, and H.~L. Miler, eds.
%   (Cambridge University Press, 2007).

% \end{thebibliography}

\end{document}